\documentclass[twocolumn, pra, showpacs,superscriptaddress]{revtex4}
\usepackage{amsfonts}
\usepackage{amsmath}
\usepackage{amssymb}
\usepackage{graphicx}

\begin{document}

\title{Maximum Entangled State in Ultracold Spin-1 Mixture}
\author{Jie Zhang}
\email{zhangjie01@tyut.edu.cn}
\affiliation{College of Physics, Taiyuan University of Technology, Taiyuan 030024,
Shanxi, People's Republic of China}
\author{Longsheng Yu}
\affiliation{College of Physics, Taiyuan University of Technology, Taiyuan 030024,
Shanxi, People's Republic of China}
\author{Zezhen He}
\affiliation{College of Physics, Taiyuan University of Technology, Taiyuan 030024,
Shanxi, People's Republic of China}
\author{Pengjun Wang}
\email{Pengjun_wang@sxu.edu.cn}
\affiliation{State Key Laboratory of Quantum Optics and Quantum Optics Devices, Institute of Opto-electronics, Shanxi University
Taiyuan Shanxi 030006, People's Republic of China}

\begin{abstract}
Inspired by the method that can deterministically generated the massive entanglement through phase transitions, we study the ground state properties of a spin-1 condensate mixture, under the premise that the heteronuclear spin-exchange collision is taken
into account. We developed a effective model to analyze the binary-coupled
two-level system and studied the ground state phase transitions. Three
representative quantum states with the same number distribution are studied
and distinguished through the number fluctuations. We demonstrate that there
will be the Greenberger-Horne-Zeilinger (GHZ) state in the mixture if the
the extra magnetic field is specifically selected or adiabatically adjusted.
One advantage of preparing entangled states in mixtures is that we only need
to adjust the external magnetic field, instead of considering the
microwaves-magnetic cooperation. Finally we estimate the feasibility of
experimentally generating the heteronuclear
many-body entanglement in the alkali-metal atomic mixture.
\end{abstract}

\maketitle

\section{introduction}

Quantum entanglement \cite{Quantum entanglement} is a fundamental yet
intriguing quantum phenomena, the coherence and non-locality make it
distinguished from classical physics. Besides the essence of quantum
mechanics, the generated entanglement states are important resources for
applications in many fields, including quantum information \cite{ref2},
quantum computation \cite{ref3,ref4} quantum metrology \cite{ref5,ref6}%
 and quantum cryptography \cite{ref7}.
How to generate stable entangled states has become a hot topic in
fundamental science frontier recent years.

Greenberger-Horne-Zeilinger (GHZ) state is one of the ultimate goals for
quantum information and quantum metrology, due to its extraordinary
precision guaranteed by fundamental quantum principles. However, it is
challenging to create entanglement states in multi-particle ensembles, although
efforts have been made at different
platforms \cite{GHZ1,GHZ2,GHZ3,GHZ4,GHZ5,GHZ6}, including 20-qubits GHZ state in Rydberg atom
arrays \cite{science2019_1} and superconducting quantum circuit \cite{science2019_2}.
The required precision of
the control and technical difficulties making it difficult to increase the
size of the entanglement states. Recently, GHZ states are created in optical cavity,
by shifting the energy of a particular Dicke state to convert unentangled
states into entangled states \cite{2021npj}, where 100-atom GHZ state can be
obtained with a fidelity at 0.92, and the one with 2000 atoms can be
achieved with a fidelity at 0.89.

The spin-exchanging collisions \cite{TLH,Ohmi,Law,Yi,MSChang,spinorfull} inside Bose-Einstein condensates (BEC) is also one of the
popular candidates to produce multiparticle
entanglement state, such as spin squeezed states \cite%
{L.M.Duan,squeezing1,squeezing2}, spin-nematic squeezing states \cite%
{nematic}, Dicke states \cite{Dicke Experiment1,Dicke Experiment2,Dicke
Experiment3}, and twinfock states \cite{indicator,twinfock,LiYouTwinfock}. Recent years, the condensate mixtures and
heteronuclear spin-exchanging collisions have also attracted a lot of
attentions \cite{mixture1,mixture2,mixture3,mixture4,mixture5,mixture6,mixture7,mixture8, KRb, KRbdroplet}. A unique
feature of the spin-1 mixture is that the inter-species spin-spin
interaction takes place over all three total spin F=0,1, and 2 channels.
Besides this, the Zeeman energy shift of hyperfine states for different
atoms are quite different in the same magnetic field, which make it possible
to select special spin-exchange channels.

In this paper, we theoretically analyse the possibility to generate the multi-particle GHZ
states in the mixture of spin-1 condensates, on the premise of adjusting the
external magnetic field adiabatically. The assumption of adiabatically
adjust magnetic field has been reliazed in the single spin-1 condensate,
where more than 1000-atom entangled twin-Fock condensate was generated
deterministically \cite{LiYouTwinfock}. The advantage of preparing entangled states in mixtures is that we only need
to adjust the external magnetic field, instead of considering the
microwaves-magnetic cooperation. Moreover, we propose near-deterministic generation, instead of dynamical. The latter
can also lead to fluctuations and degradation of the entanglement.

The outline of this paper is as follows: Sec.~\ref{sec:Ham} introduces the
model hamiltonian, the spin-exchange processes, the Zeeman effects, and the reduced spin-exchange process in the
extra fields. In Sec.~\ref{sec:theory}, we given a effective reduced
hamiltonian to analytically explain the special ground states and calculate the observables,
such as number fluctuation and number distribution, as well as the quantities that can characterize and measure
entanglement. In Sec.~\ref{sec:experiment}, we discuss possible
parameters ranges to capture or realize the entanglement state
experimentally. Conclusions are made at the end in Sec.~\ref{sec:con}.

\section{The model Hamiltonian}

\label{sec:Ham}

In the spinor mixture, the angular momentum coupling between heteronuclear
atoms do not obey the identity principle any more, which cause the
heteronuclear spin interaction between spin-1 atoms takes place over all
three total spin F=0,1, and 2 channels. The $inter-species$ spin-exchange
interaction can be expressed as,
\begin{equation}
V_{12}(\mathbf{r})=\frac{1}{2}(\alpha +\beta \mathbf{f}_{1}\cdot \mathbf{f}%
_{2}+\gamma P_{0})\delta (\mathbf{r}),
\end{equation}%
where the parameters $\alpha =(g_{1}+g_{2})/2,\beta =(g_{1}-g_{2})/2$, and $%
\gamma =(2g_{0}+g_{2}-3g_{1})/2,$ are the linear combinations of the
coupling constants $g_{F}=\frac{2\pi \hbar ^{2}a_{F}}{\mu }$. $a_{F}$ are $s$%
-wave scattering lengths in the total spin $F$=0,1,2 channels. $\mu =\frac{%
\mu _{1}\mu _{2}}{\mu _{1}+\mu _{2}}$ denotes the reduced mass for atoms in
different species and $\hbar$ is the Planck's constant. $P_{0}$ projects an
inter-species pair into spin singlet state only through the F=0 channel.

Including the $intra-species$ spin-exchange interaction, the spin-dependent Hamiltonian for the
mixture under the single-mode approximation (SMA) \cite{Law,Yi} finally
takes the form $H=H_{1}+H_{2}+H_{12}$, where
\begin{eqnarray}
H_{1} &=&c_{1}\beta _{1}\frac{\mathbf{L}_{1}^{2}}{N_{1}}+p_{1}(\hat{n}_{1}-%
\hat{n}_{-1})+q_{1}(\hat{n}_{1}+\hat{n}_{-1}), \\
H_{2} &=&c_{2}\beta _{2}\frac{\mathbf{L}_{2}^{2}}{N_{2}}+p_{2}(\hat{n}%
_{1}^{\prime }-\hat{n}_{-1}^{\prime })+q_{2}(\hat{n}_{1}^{\prime }+\hat{n}%
_{-1}^{\prime }),\\
H_{12} &=&c_{12}\beta \frac{\mathbf{L}_{1}\cdot \mathbf{L}_{2}}{N_{12}}%
+c_{12}\gamma \frac{\Theta _{12}^{\dag }\Theta _{12}}{N_{12}}.
\label{Ham12}
\end{eqnarray}
The number operators for two species are defined as
\begin{align}
\hat{n}_{0}& =\hat{a}_{0}^{\dag }\hat{a}_{0},\hat{n}_{\pm 1}=\hat{a}_{\pm
1}^{\dag }\hat{a}_{\pm 1}, \\
\hat{n}_{0}^{\prime }& =\hat{b}_{0}^{\dag }\hat{b}_{0},\hat{n}_{\pm
1}^{\prime }=\hat{b}_{\pm 1}^{\dag }\hat{b}_{\pm 1},
\end{align}%
and $p_{1,2}(q_{1,2})$ denotes the linear (quadratic) Zeeman field energy
for the two species.
Parameter $\beta _{1,2}=4\pi\hbar ^{2}a_{1,2}/\mu _{1,2}$ denotes the
intra-species spin-exchange interaction, and the sign of which determines
the polar (+) or ferromagnetic (-) feature of a spinor condensate
respectively.  $c_{1}$, $c_{2}$ and $c_{12}$ are tunable
parameters determined by particle numbers and trapping frequencies.
$\mathbf{L}_{1}=a_{m_{1}}^{\dag }(\mathbf{f}_{1})_{m_{1}n_{1}}a_{n_{1}}$ and
$\mathbf{L}_{2}=b_{m_{2}}^{\dag }(\mathbf{f}_{2})_{m_{2}n_{2}}b_{n_{2}}$ are
the spin-1 angular momentum operator for the two species, operator $
\Theta _{12}^{\dag }$ creates a singlet pair with one atom each from the two
species. More comprehensive and detailed presentations given by $
a_{m_{1}}(b_{m_{2}})$ and $a_{m_{1}}^{\dag}$($b_{m_{2}}^{\dag}$) are
presented in the appendix ~\ref{sec:detail-1}.

\section{Ground state properties}
\label{sec:theory}

To study the effects of the interspecies spin-exchange
interaction, one would like to suppress the
intraspecies parts. This can be done by choosing
suitable parameters $q_{1,2}$. A microwave \cite{twinfock} can also contribute to the parameter $q_{1,2}$,
so that the linear and the quadratic Zeeman field can actually be controlled
independently. Consider the situation that linear Zeeman field is comparable
to the quadratic Zeeman field, the case $p_{1}=-q_{1}$ can suppress the $%
\hat{n}_{1}$ population of the first species. Similarly, if $p_{2}=q_{2} $,
the population on the $\hat{n}_{-1}^{\prime}$ of the second species are
suppressed. In this paper we mainly focus on the process-4 (see appendix ~\ref{sec:detail-1}) which has the
advantage to discuss the heteronuclear spin-exchange physics in the subspace
with the total magnetization equals to zero.

\subsection{Effective Hamiltonian and basis}

In the chosen and fixed extra fields, the total Hamiltonian (\ref{Ham12})
reduces to a more simple and symmetric case with only four Zeeman components
involved:
\begin{align}
H_{12}& =-[\Gamma _{1}+\Gamma _{2}](\hat{n}_{-1}\hat{n}_{1}^{\prime}+\hat{n}%
_{0}\hat{n}_{0}^{\prime })-\Gamma _{2}\hat{n}_{-1}\hat{n}_{1}^{\prime}
\notag \\
& +\Gamma _{1}(\hat{a}_{-1}^{\dag }\hat{b}_{1}^{\dag }\hat{b}_{0}\hat{a}_{0}+%
\hat{a}_{0}^{\dag }\hat{b}_{0}^{\dag }\hat{a}_{-1}\hat{b}_{1}),
\label{Ham12-2}
\end{align}
where
\begin{equation}
\Gamma _{1}=c_{12}(g_{1}-g_{0}),\Gamma _{2}=c_{12}\frac{g_{2}-g_{1}}{2}.
\end{equation}
If $g_{2}$ is independently tunable, we can achieve a special point with $%
\Gamma _{2}=0,$ which make the Hamiltonian (\ref{Ham12-2}) more symmetric
and analyzable. This is a binary-coupled two-level system, each species can
be consider as a two-level quantum oscillator. Similar to the two-site
Bose-Hubbard model, terms like $\hat{a}_{-1}^{\dag }\hat{b}_{1}^{\dag }\hat{b%
}_{0}\hat{a}_{0}$ can denote the "hopping-like" interactions, while the term
$\hat{n}_{-1}\hat{n}_{1}^{\prime}$ or $\hat{n}_{-1}\hat{n}_{1}^{\prime}$
represents the density-interaction between particles. The Hilbert space can
be expanded by the Fock states that describe the particle-occupation on four
different energy levels $\left\vert n_{-1},n_{0}\right\rangle \otimes
\left\vert n_{0}^{\prime },n_{1}^{\prime }\right\rangle ,$ or alternatively,
we can rewrite them as $\left\vert m_{1}\right\rangle \otimes \left\vert
m_{2}\right\rangle $, with $m_{1}=n_{-1}-n_{0}$, and $m_{2}=n_{0}^{\prime }-n_{1}^{\prime }$. If
we consider the special case of $N_{1}=N_{2}=N$ and for even $N$, we can
rewrite the basis as:
\begin{equation}
\left\vert m_{1}\right\rangle \otimes \left\vert m_{2}\right\rangle
=\left\vert M,m\right\rangle ,  \label{m1m2}
\end{equation}
where we define $M=m_{1}+m_{2}$ and $m=m_{1}-m_{2}$. Still further, the
unique properties of process-4 (vanishing the total spin) guarantees that $%
M=0$, and only one parameter $m$ is needed to describe the different basis:
\begin{equation}
\left\vert n_{-1},n_{0},n_{0}^{\prime },n_{1}^{\prime }\right\rangle
\rightleftharpoons \left\vert m\right\rangle .
\end{equation}%
The ground state is a superposition states of all basis, $\left\vert
G\right\rangle =\sum\nolimits_{m}\Psi _{m}\left\vert m\right\rangle $.

\begin{figure}[t]
\includegraphics[width=3.5in]{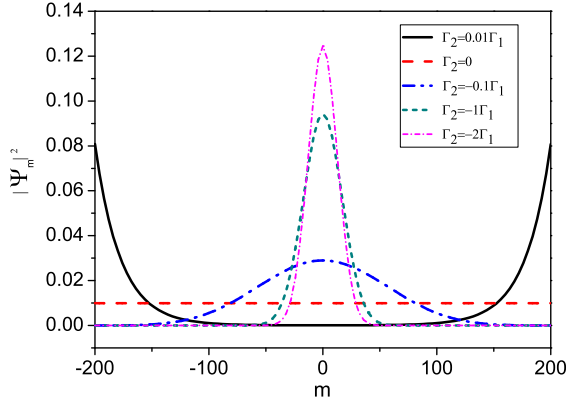}
\caption{(Color online) The ground state amplitude distribution $\left\vert \Psi
_{m}\right\vert ^{2}$ of spin-1 mixture in a fixed magnetic field. The
values of m represent different basis, and the parameter $\Gamma_{2}$ is
assumed to be adjustable. The total numbers are $N_{1}=N_{2}=100$.}
\label{mm}
\end{figure}

\subsection{Different phases}

For analysis purposes, we first consider a hypothetical Hamiltonian which is
more symmetrical and analyzable:
\begin{equation}
H=\Gamma _{1}(\hat{A}^{\dag }\hat{B}+\hat{B}^{\dag }\hat{A})-(\Gamma
_{1}+\Gamma _{2})(\hat{n}_{-1}\hat{n}_{1}^{\prime }+\hat{n}_{0}\hat{n}%
_{0}^{\prime }).  \label{Ham12-3}
\end{equation}
The parameter $\Gamma _{2}$ is adjustable which can range from positive to
negative. Without loss of generality, we first consider the positive
parameter $\Gamma _{1}$ and fixed it to be one. $\hat{A}^{\dag }=\hat{a}%
_{0}^{\dag}\hat{b}_{0}^{\dag}$, and $\hat{B}^{\dag }=\hat{a}_{-1}^{\dag }%
\hat{b}_{1}^{\dag }$ are the pair creation operators for each species.
Minimizing the energy (\ref{Ham12-3}), one can get the profile $\left\vert
\Psi _{m}\right\vert ^{2}$ for different phases. The numerical results are
demonstrated in Fig.\ref{mm}, where the total particle numbers are $%
N_{1}=N_{2}=100$. It's obviously that the special point with $\Gamma _{2}=0$
are the critical point.
\begin{figure}[tbp]
\includegraphics[width=3.5in]{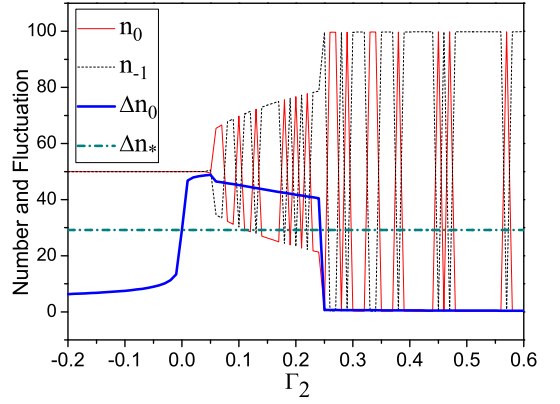}
\caption{(Color online) The Zeeman-level number distributions $n_{0}$(red
solid) and $n_{-1}$(black dashed), the number fluctuation $\bigtriangleup
n_{0}$(blue solid) of the species A are presented, as the interaction
parameter $\Gamma _{2}$ is adjustable and in the unit of $\Gamma_{1}$. The
total numbers are $N_{1}=N_{2}=100$. The black dashed line ($\bigtriangleup
n_{\ast }$) which denote the 0-component number fluctuation of the special
ground state $\left\vert G\right\rangle _{\ast}$ is also displayed for
comparison. The Zeeman-level number distributions and fluctuations of the
species B which are either strictly equal to or opposite to the
presentations, are not shown.}
\label{N0N1}
\end{figure}

\subsubsection{special states when $\Gamma _{2}<0$}

When $\Gamma _{2}<0$, the ground state profile $\left\vert \Psi
_{m}\right\vert ^{2}$ presents a Gaussian distribution. In this region, the
special point with $\Gamma _{2}=-\Gamma _{1}$ is particularly
comprehensible, that means there are no density interactions between
particles and the hopping process $\hat{A}^{\dag }\hat{B}+\hat{B}^{\dag }%
\hat{A}$ dominates the system. The ground state is a direct product of two
independent coherent states,
\begin{align}
\left\vert G\right\rangle _{c} =\frac{(\hat{a}_{0}^{\dag }+\hat{a}%
_{-1}^{\dag })^{N}\left\vert 0\right\rangle }{\sqrt{2^{N}N!}}\otimes \frac{(%
\hat{b}_{1}^{\dag }+\hat{b}_{0}^{\dag })^{N}\left\vert 0\right\rangle }{%
\sqrt{2^{N}N!}}  \notag \\
=\frac{1}{2^{N}N!}(\hat{a}_{0}^{\dag }\hat{b}_{1}^{\dag }+\hat{a}_{0}^{\dag }%
\hat{b}_{0}^{\dag }+\hat{a}_{-1}^{\dag }\hat{b}_{1}^{\dag }+\hat{a}%
_{-1}^{\dag }\hat{b}_{0}^{\dag })^{N}\left\vert 0\right\rangle.
\end{align}
The average numbers of atoms in the four components are exactly all equal to
$N/2$, the number fluctuations on this point is exactly
\begin{equation}
\triangle n_{-1}=\triangle n_{0}=\triangle n_{0}^{\prime }=\triangle
n_{1}^{\prime }=\sum\nolimits_{m}m^{2}\Psi _{m}^{2}.
\end{equation}
When $\Gamma_{2}<-\Gamma_{1}$, the coherence of product state still
maintain, but can be shifted as $\Gamma_{2}$ is adjusted. The decreasing of $%
\Gamma_{2}$ can suppress particle fluctuations between different components
and narrow the Gaussian toward the delta-function \cite{Hocat}, which
corresponds to a Fock state,
\begin{equation}
\left\vert m=0\right\rangle=\left\vert\frac{N}{2},\frac{N}{2},\frac{N}{2},%
\frac{N}{2}\right\rangle.
\end{equation}
In the region $-\Gamma _{1}<\Gamma _{2}<0$, as demonstrated in Fig.\ref{mm}, the
Gaussian distribution is broaden as $\Gamma _{2}$ is increasing, unit the
distribution is totally flat on the point $\Gamma _{2}=0$.

When $\Gamma_{2}=0$, this point gives us a uniform distribution state with $%
\left\vert \Psi _{m}\right\vert ^{2}=\frac{1}{N+1}$, or, $\left\vert
G\right\rangle _{\ast }=\sum\nolimits_{m}\frac{1}{\sqrt{N+1}}\left\vert
m\right\rangle$. Equivalently, we can proof that this state has a analytical
form:
\begin{align}
\left\vert G\right\rangle _{\ast}& =\frac{1}{2^{N}N!}(\hat{a}_{0}^{\dag }%
\hat{b}_{0}^{\dag }+\hat{a}_{-1}^{\dag }\hat{b}_{1}^{\dag })^{N}\left\vert
0\right\rangle.
\end{align}
The number fluctuations on this point is exactly,
\begin{equation}
\triangle n_{0,-1}=\triangle n_{0,1}^{\prime}=\sqrt{\frac{N(2N+1)}{6}-\frac{%
N^{2}}{4}}.
\end{equation}

\subsubsection{special state when $\Gamma _{2}>0$}

\begin{figure}[h]
\includegraphics[width=3.5in]{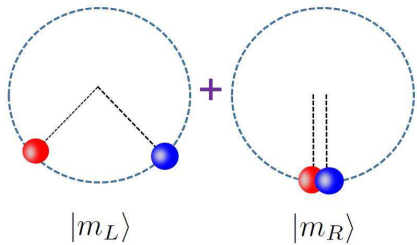}
\caption{(Color online) Two-particle Newton's pendulum can demonstrate the
characteristics of the maximum entanglement state in spin-1 BEC mixture. Red
balls denote the particles from A system, and blue balls denote the
particles from B system. Two balls on the equilibrium position can be
regarded as the dead-cat state or $\left\vert m_{R}\right\rangle$, while two
balls on the highest position is the live-cat state or $\left\vert
m_{L}\right\rangle$.}
\label{pendulum}
\end{figure}

When $\Gamma _{2}>0$, the ground state split into a bimodal distribution
state, which indicates a Schrodinger Cat like state (CAT-state) \cite{Hocat}: $\left\vert
G\right\rangle _{cat}=\frac{1}{\sqrt{2}}(\left\vert m_{L}\right\rangle
+\left\vert m_{R}\right\rangle)$,
\begin{eqnarray}
\left\vert m_{L}\right\rangle &=&\left\vert m=-2N\right\rangle =\left\vert
N,0,0,N\right\rangle , \\
\left\vert m_{R}\right\rangle &=&\left\vert m=+2N\right\rangle =\left\vert
0,N,N,0\right\rangle ,
\end{eqnarray}%
or simply denoted as,
\begin{equation}
\left\vert G\right\rangle _{cat}=\frac{1}{\sqrt{2}}(\left\vert
-1\right\rangle _{1}^{\otimes N}\left\vert 1\right\rangle _{2}^{\otimes
N}+\left\vert 0\right\rangle _{1}^{\otimes N}\left\vert 0\right\rangle
_{2}^{\otimes N}),
\end{equation}
which is illustrated by a two-body Newton's pendulum in Fig.\ref{pendulum}.

The splitting from Gaussian distribution to CAT-state distribution is in
fact a spontaneous symmetry breaking process. The latter is extremely
unstable and can exist only when $\Gamma _{2}$ is slightly greater than 0.
The comparable strength between "hopping" and "interaction" can lead to the
CAT-state, otherwise, the system has to chosen one possible state at a time
between two Fock states $\left\vert m_{L}\right\rangle$ and $%
\left\vert m_{R}\right\rangle$. We calculate the particle numbers and number
fluctuations on CAT-state and find that the average numbers of atoms in the
four components are exactly all equal to $N/2$, while the number
fluctuations are macroscopical ($\sim N$), which indicate that this is a
fragmented BEC mixture \cite{HoYip}.

It is difficult to distinguish $\left\vert G\right\rangle _{\ast}$ from $%
\left\vert G\right\rangle _{cat}$ by the number distributions ($%
n_{0,-1}=n_{0,1}^{\prime}=N/2$), however, the number fluctuations can reveal
the secret and distinct different phases, see Fig.\ref{N0N1}. When $\Gamma
_{2}<0 $, the fluctuation is small and flat, then the fluctuation of
particle number increases significantly near $\Gamma _{2}$=0, that means
system enter the state $\left\vert G\right\rangle _{\ast}$. In the region $%
0<\Gamma _{2}<\frac{1}{4}$ where $\Gamma _{2}$ is tinily greater than 0, we
notice that the number fluctuations of state $\left\vert G\right\rangle
_{cat}$ is higher that the number fluctuations of the state $\left\vert
G\right\rangle_{\ast}$. Moreover, two number states are populated
simultaneously in this region, that means there are multi-particle GHZ
states in the spinor mixture.

When $\Gamma_{2}>\frac{1}{4}$, the number fluctuation drop dramatically and
the number distribution of $n_{0}$ is capricious between 0 and 100 as $%
\Gamma_{2}$ is adjusted. This reflects the fact that there is a trivial Fock
state after the spontaneous symmetry breaking, however, which number state
to populate is totally random.

\subsection{The GHZ state and entanglement entropy}

The GHZ state is a certain type of entangled state that involves at least
three qubits, and its most remarkable characteristics can be demonstrated
intuitively by the density matrix. For example, one maximally entangled
state of tripartite, $\Phi =\frac{1}{\sqrt{2}}(\left\vert 000\right\rangle
+\left\vert 111\right\rangle ),$ its density operator is
\begin{equation*}
\rho =\frac{1}{2}\left[
\begin{array}{cccccccc}
1 & 0 & 0 & 0 & 0 & 0 & 0 & 1 \\
0 & 0 & 0 & 0 & 0 & 0 & 0 & 0 \\
0 & 0 & 0 & 0 & 0 & 0 & 0 & 0 \\
0 & 0 & 0 & 0 & 0 & 0 & 0 & 0 \\
0 & 0 & 0 & 0 & 0 & 0 & 0 & 0 \\
0 & 0 & 0 & 0 & 0 & 0 & 0 & 0 \\
0 & 0 & 0 & 0 & 0 & 0 & 0 & 0 \\
1 & 0 & 0 & 0 & 0 & 0 & 0 & 1%
\end{array}%
\right] ,
\end{equation*}
which is a square matrix with non-zero elements in the four corners and zero
elements elsewhere. The GHZ state in our Fock state basis can be described
as $\Phi_{GHZ} =\frac{1}{\sqrt{2}}(\left\vert m_{L}\right\rangle +\left\vert
m_{R}\right\rangle ),$ and its density operator is

\begin{equation*}
\rho =\frac{1}{2}\left[
\begin{array}{ccccc}
1 & 0 & ...... & 0 & 1 \\
0 & ... & ...... & ... & 0 \\
\begin{array}{c}
. \\
. \\
.%
\end{array}
& ... & ...... & ... &
\begin{array}{c}
. \\
. \\
.%
\end{array}
\\
0 & ... & ...... & ... & 0 \\
1 & 0 & ...... & 0 & 1%
\end{array}%
\right] ,
\end{equation*}
with the dimension of matrix equals to the numbers of basis (\ref{m1m2}).

For the bipartite quantum system, the measure of entanglement can be given
by the partial von Neumann entropy. The von Neumann entropy of a state is
defined as
\begin{equation}
S(\rho)=-Tr(\rho\log{\rho}),
\end{equation}
where the symbol $\rho$ is the density operator for the system and Tr(...)
denotes the trace operation. For the bipartite A-B system, the degree of
entanglement can be defined as the von Neumann entropy of reduced density
operators $\rho_{A} $ or $\rho_{B}$, which is defined as ,
\begin{equation}
\rho_A=-Tr_B[\rho_{AB}],\rho_B=-Tr_A[\rho_{AB}],
\end{equation}
where $\rho_{AB}$ is the density operator of the whole system. We can
diagonalize $\rho_A$( or $\rho_B$) so that the partial von Neumann entropy for the
entanglement degree can be writhen as
\begin{equation}
S(\rho_A)=-\sum_n(\rho_n\log{\rho_n}),  \label{entropy}
\end{equation}
where the symbols $\rho_n$ are the non-negative eigenvalues of the reduced
density operator $\rho_{A}$, and sum to unity, $\sum{\rho_{n}}=1$.
\begin{figure}[tbp]
\includegraphics[width=3.5in]{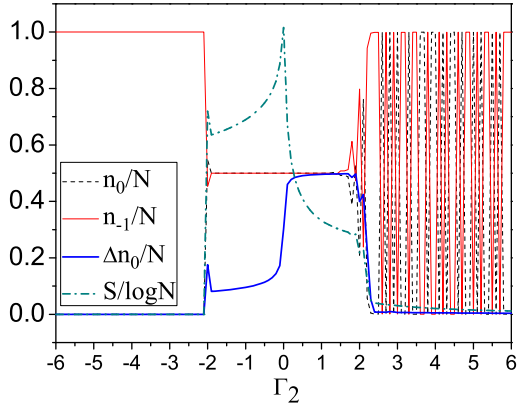}
\caption{(Color online) The Zeeman-level number distributions $n_{0}$ (black
dashed) and $n_{-1}$ (red solid), the number fluctuation $\bigtriangleup
n_{0}$ (blue solid), and the partial von Neumann entropy S (green dash-dot)
of special A are presented, as the interaction parameter $\Gamma_{2}$ is
adjustable and in the unit of $\Gamma_{1}$. The number distributions and
fluctuations of particles in special B which are either strictly equal to or
opposite to the presentations are not shown. The total numbers are $%
N_{1}=N_{2}=20$.}
\label{entr}
\end{figure}

In order to understand the properties more comprehensively, we consider the
full-space expansion with the basis $\left\vert m_{1}\right\rangle \otimes
\left\vert m_{2}\right\rangle$, and $N=N_{1}=N_{2}=20$. We diagonalize the Hamiltonian (\ref{Ham12-2}) and
track the entanglement characteristic using the partial von Neumann entropy S($\rho_{A}$),
whiich is illustrated in different interacting regions, see Fig.(\ref%
{entr}). We found that there is a zone with no-zero entropy (no-matter $%
\Gamma_{2}<0$ or $\Gamma_{2}>0$), which is always associated with the
equally distributed particles on the four Zeeman-level:
\begin{equation}
n_{-1}=n_{0}=n_{0}^{\prime}=n_{1}^{\prime}=N/2.  \label{equal}
\end{equation}
In the $\Gamma_{2}<0$ region, the degree of entanglement is higher and keep
growing until $\Gamma_{2}=0$. On the $\Gamma_{2}=0$ point, the top value $%
S(\rho_A)$=1, confirms the knowledge that the uniform distribution state $%
\left\vert G\right\rangle _{\ast}$ is the maximally entangled State.

\begin{figure}[tbp]t
\includegraphics[width=3.5in]{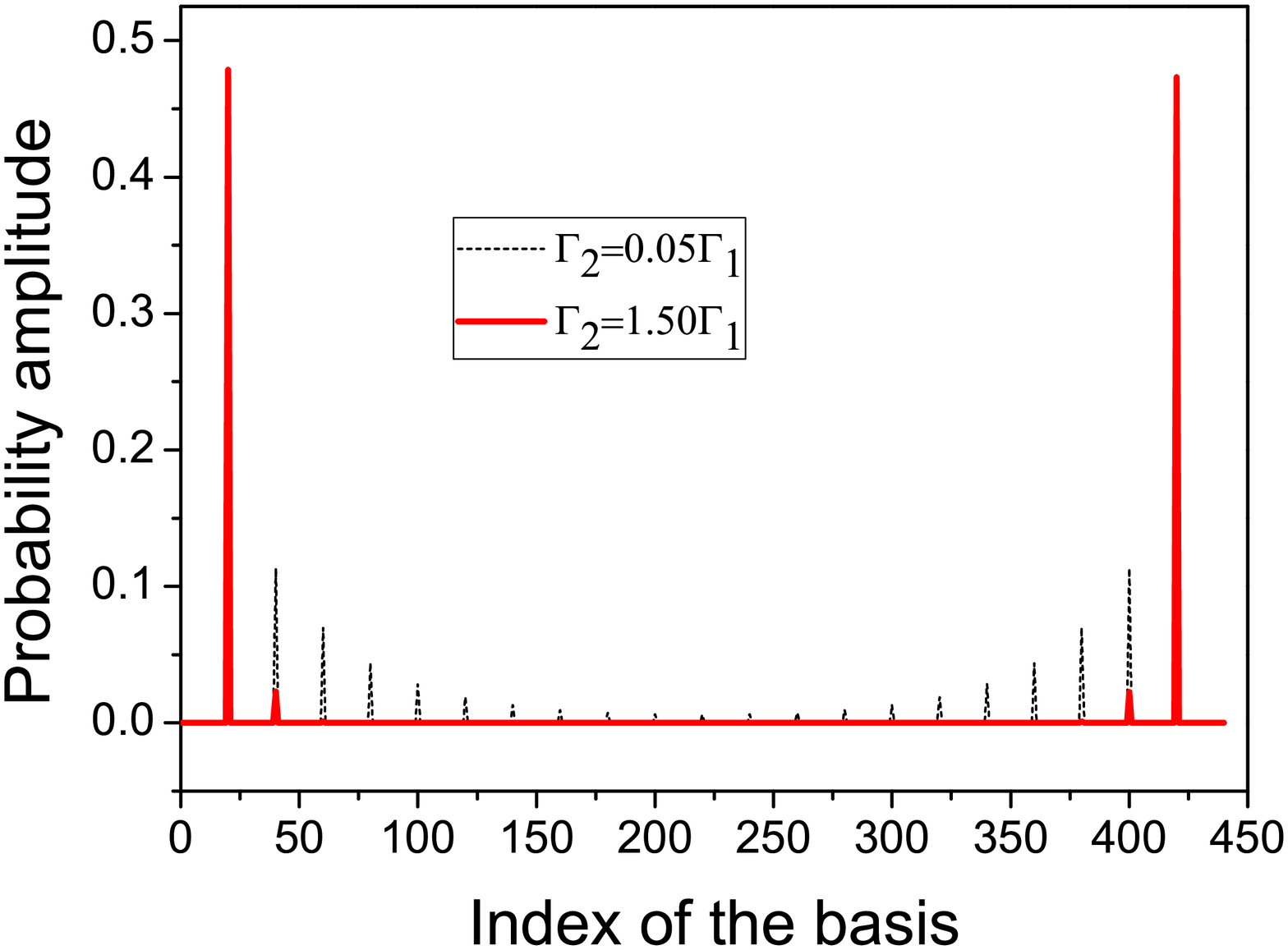}
\caption{(Color online) The ground state amplitude distribution $\left\vert \Psi
_{M,m}\right\vert ^{2}$ of spin-1 mixture. The horizontal axis displays the index of
basis and the ordinate value is the amplitude. We pick up two representative points,
$\Gamma_{2}=0.05\Gamma_{1}$ and $\Gamma_{2}=1.5\Gamma_{1}$, to illustrate the basic properties.
The total particle number is $N_{1}=N_{2}=20$ and the total number of basis is 441.}
\label{amp}
\end{figure}

So far, we have discussed the entanglement properties of the coherent state,
the uniform distribution state, and the CAT-state. Since particles are
indistinguishable in BECs and it is not possible to determine which
particles occupy a particular spatial pattern, one should explore other
possible ways to divide the system. In consideration of the fact that how
many particles occupy a spatial pattern can be measured physically, the
subspace A or B can be divided according to two distinguishable spatial
patterns. In our case, we divided the system by the two spatially
incompatible BECs, and each BEC contains two models respectively. Even if
there are 40 particles in the system, the system's entanglement is
still based on only four modes. In this perspective, the maximum entangled
state of the four-mode BEC-mixture should be the uniform distribution state $%
\left\vert G\right\rangle_{\ast}$. The well-known GHZ or Cat state, on the
contrary, present the lower entanglement degree. In Fig.(\ref{amp}), we
illustrated the ground state profile $\left\vert\Psi _{M,m}\right\vert ^{2}$s
for two representative points in region $\Gamma_{2}>0$, and the red line confirms the
Schroedinger's Cat properties: simultaneous occupation on two opposite states,
large fluctuations and no zero entanglement entropy.

\begin{figure}[tbp]
\includegraphics[width=3.3in]{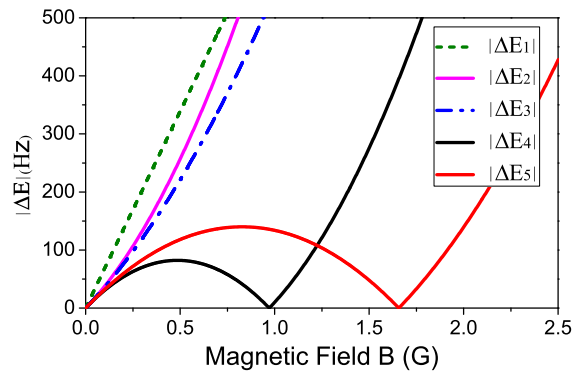}
\caption{(Color online) The dependence of heteronuclear magnetic energy
differences on the applied magnetic field. Two processes represented by $%
\Delta E_{4}$ and $\Delta E_{5}$ have zero-crossing points at $B=0.97G$ and $%
B=1.69G$ respectively, while other processes show large detuning. The energy
differences $\Delta E_{i}$ can be both positive and negative, here we took
their absolute values to make the picture more compact.}
\label{Mag}
\end{figure}

\begin{figure}[tbp]
\includegraphics[width=3.3in]{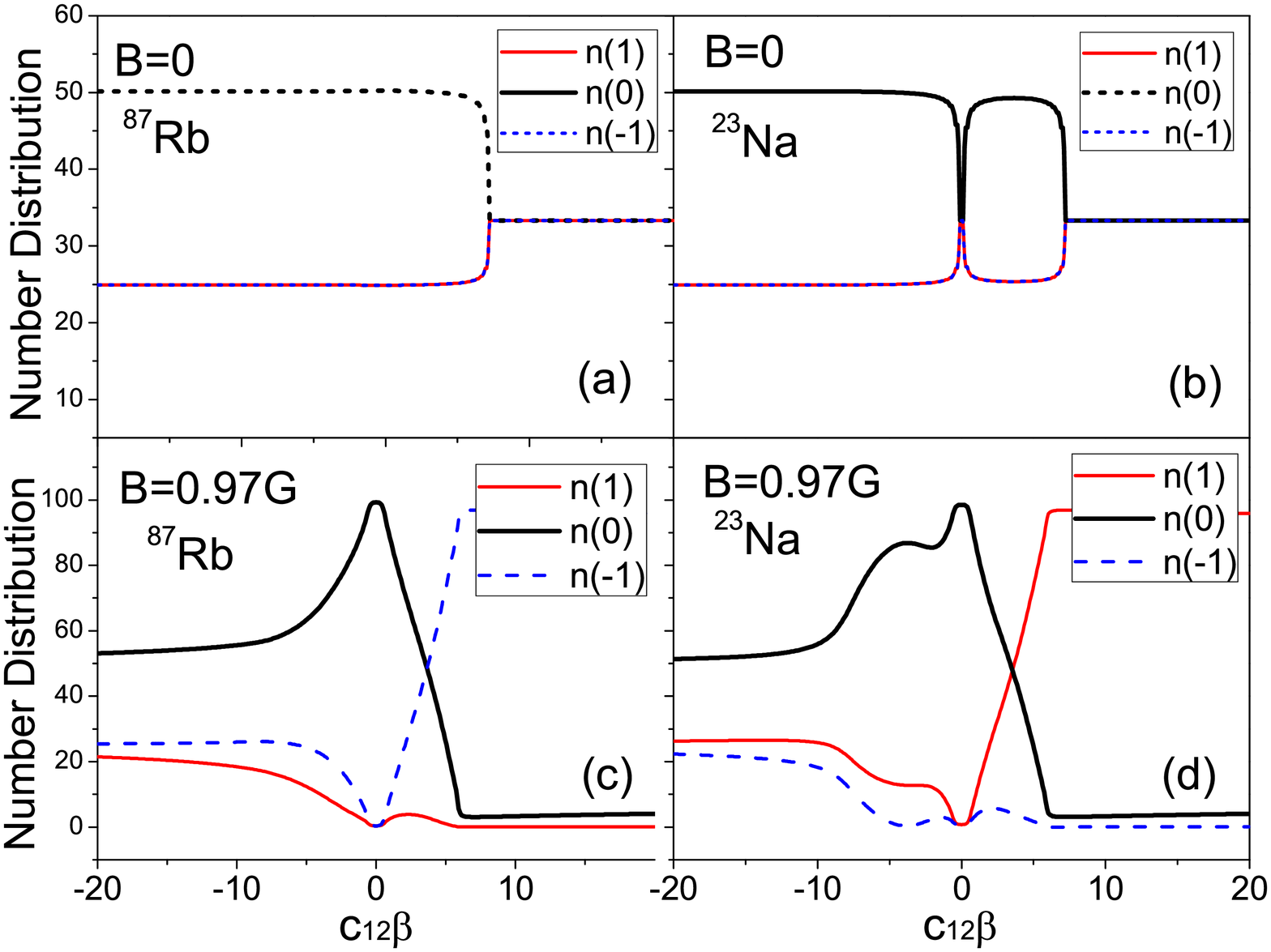}
\caption{(Color online) Numerical calculation of Zeeman level number
distributions of Na-Rb mixture in a fixed magnetic field $B_{0}=0$ and $%
B_{0}=0.97G$. The intraspecies interactions of Na species are fixed to be $%
c_{1} \protect\beta _{1}=1$, and other interacting parameters are in the
unit of $c_{1} \protect\beta _{1}$. The intraspecies interactions of Rb
species is set to be $c_{2}\protect\beta _{2}=-2$ without loss of
generality. The interspecies interaction $c_{12}\protect\beta $ is assumed
to be adjustable and can range from negative to positive. parameter $\protect\gamma $ is much smaller than $\protect\beta $ and
has been neglected.}
\label{Num}
\end{figure}

\section{Parameter estimation and experimental realization}

\label{sec:experiment}

In this section, we discuss the possibility of realizing the inter-species
entanglement state in the ultra-cold Na$^{23}$ and Rb$^{87}$ mixture. The
Zeeman shifts $E_{m_{1}}^{(1)}$ and $E_{m_{2}}^{(2)}$ can be calculated by
the Breit-Rabi formula, which can give the corresponding Zeeman energy
differences $\Delta E_{i}$ (before and after scatting) for the 5 processes:
\begin{eqnarray*}
\Delta E_{1} &=&E_{0}^{(1)}+E_{-1}^{(2)}-E_{-1}^{(1)}-E_{0}^{(2)}, \\
\Delta E_{2} &=&E_{0}^{(1)}+E_{1}^{(2)}-E_{1}^{(1)}-E_{0}^{(2)}, \\
\Delta E_{3} &=&E_{0}^{(1)}+E_{0}^{(2)}-E_{-1}^{(1)}-E_{1}^{(2)}, \\
\Delta E_{4} &=&E_{0}^{(1)}+E_{0}^{(2)}-E_{1}^{(1)}-E_{-1}^{(2)}, \\
\Delta E_{5} &=&E_{-1}^{(1)}+E_{1}^{(2)}-E_{1}^{(1)}-E_{-1}^{(2)}.
\end{eqnarray*}
By appropriately applying the Zeeman field \cite{mixture8}, people can
select out one special spin-exchange process, with the related Zeeman energy
differences $\Delta E_{i}$ vanished. The other
spin-exchange channels are forbidden, because the corresponding Zeeman
energy differences show large detunings, see Fig.(\ref{Mag}).

In Fig.(\ref{Num}), we shows the dependence of the number distributions on
the inter-species interaction $c_{12}\beta$ at two fixed values of magnetic
field. The numerical results are obtained through the approach of exact
diagonalizing the total Hamiltonian $H=H_{1}+H_{2}+H_{12}$, and $
H_{12}$ is the original Hamiltonian (\ref{A6}). The basis we
considered is the direct product of spin-1 Fock states:
\begin{equation}
\left\vert n_{1}^{(1)},n_{0}^{(1)},n_{-1}^{(1)}\right\rangle \otimes
\left\vert n_{1}^{(2)},n_{0}^{(2)},n_{-1}^{(2)}\right\rangle,
\end{equation}
but, the total magnetization is limited with $m_{tot}=m_{1}+m_{2}=0$.

When the magnetic field is absent, there are FF phase (or coherent state)
when $c_{12}\beta \ll -1$, and the AA phase when $c_{12}\beta \gg 1$
\cite{mixture3}, the later is a fragmented state with particle number
equally distributed in three Zeeman levels. The $c_{12}\beta=0$ point means independent,
which shows the characteristic particle number distribution $(N/4,N/2,N/4)$ and $%
(N/3,N/3,N/3)$ for Rb$^{87}$ and Na$^{23}$ respectively.

On the point $c_{12}\beta =0$ and B=0.97G, particles only distribute on the $n_{0}^{(1)}$ and $%
n_{0}^{(2)}$ components respectively, that is the typical quadratic Zeeman
effect \cite{LiYouTwinfock,n396,n443} for spin-1 BEC.

When the inter-species
spin-exchange interaction is not zero, and $c_{12}\beta >0$.
The remarkable properties for the spin-exchange effects are
highlighted, as illustrated in Fig.2(c) and Fig.2(d). The selected process-4
will finally drive the the ground state number distributions from $(0,N,0)$
to $(0,0,N)$ for Rb BEC, and from $(0,N,0)$ to $(N,0,0)$ for Na BEC. We
notice that there is a crossover point, where the numbers are equally
distributed on four Zeeman-levels. That means the interacting region near
the crossover point can possibly contain the three distinctive states that
we have discussed before.

The background s-wave scattering lengths of Na$^{23}$ and Rb$^{87}$
heteronuclear interaction have been calculated in \cite{mixture7}. They are $%
a_{0}$=82.71$a_{B}$, $a_{1}$=81.4$a_{B}$,and $a_{2}$=78.9$a_{B}$, which
results in the reality that
\begin{equation}
\Gamma _{1}=\frac{2\pi\hbar ^{2}a_{B}}{\mu}(-1.31),\Gamma_{2}=\frac{2\pi\hbar ^{2}a_{B}}{\mu}(-2.5),
\end{equation}
and we have $\Gamma_{2}$:$\Gamma_{1}$$\approx1.9$. The ground state amplitude distribution calculated under
the above parameters, is quite similar to the
result illustrated in Fig.(\ref{amp}), and the partial von Neumann entropy is about $S(\rho_A)$=0.29924.

\section{Conclusion}

\label{sec:con} To conclude, we study the quantum ground states of the spin-1
mixture,  where two adjacently prepared ultra-cold atomic gases can mapping to the double-well model, and still further, there are two hyperfine states in each well. Firstly, we studied three distinctive ground states, they are the mixed coherent state $\left\vert G\right\rangle _{c}$,
the uniform distribution state $\left\vert G\right\rangle _{\ast}$, and the CAT-state $\left\vert G\right\rangle _{cat}$. We distinguished them through number fluctuations on different Zemman-levels, and gave a analytical expression of state $\left\vert G\right\rangle_{\ast}$.
Furthermore, we calculated the partial von Neumann entropy in different interacting region, and found that both $\left\vert G\right\rangle _{c}$ and $\left\vert G\right\rangle _{cat}$ had non zero entanglement entropy. The von Neumann entropy for $\left\vert G\right\rangle{\ast}$ is exactly equals to 1, which indicated that this is a maximum mixed state.
The ground state amplitude distribution for $\left\vert G\right\rangle _{c}$ is gaussian, for $\left\vert G\right\rangle _{cat}$, it is bimodal. The state $\left\vert G\right\rangle_{\ast}$, however, has a exactly uniform ground state amplitude distribution.

Since particles are indistinguishable in BECs, we have to divide the system by the spatially incompatible pattern, so called mode-division. In our case, two BECs next to each other is a natural two-mode structure, more than that, two hyperfine states inside each BECs finally lead to four independent modes. In this perspective, the maximum entangled state of the four-mode BEC-mixture is the
$\left\vert G\right\rangle _{\ast}$, with entropy $S(\rho_A)$=1. The well-known GHZ state or maximum entanglement state of qubits, on the contrary, present the lower entanglement degree. The existence of the GHZ state is confirmed through the ground state amplitude distribution and the density matrix directly. Additionally, our GHZ state is not strictly meet the traditional definition, because there are four internal spin states involved simultaneously, this can be described by a Schrodinger cat like Newton-pendulum.

Finally, we consider the experimental realization. The Zeeman shifts $E_{m_{1}}^{(1)}$ and $E_{m_{2}}^{(2)}$ at the intensity about 0.97G can lead to the binary-coupled
two-level spin-exchange model, no additional necessary of applying the microwave. According to the calculation of the Na$^{23}$- Rb$^{87}$ background s-wave scattering lengths, the interaction coincidentally located in the Cat-state region. We assume that adiabatically adjusting the magnetic field can drive the mixture into a Heteronuclear GHZ state. Our results highlight the significant promises for experimental generation of GHZ state in atomic condensate simply through adjusting the magnetic field.

This work is supported by the Natural Science Foundation of Shanxi Province
(Grant No. 202103021224051)

\appendix

\section{The full Hamiltonian and basis}

\label{sec:detail-1} We take the first species for example, where $\mathbf{L}%
_{1}=\hat{a}_{m_{1}}^{\dag }(\mathbf{f}_{1})_{m_{1}n_{1}}\hat{a}_{n_{1}}$
are defined in terms of $3\times 3$ spin-1 matrices $f_{x},$ $f_{y},$and $%
f_{z}$, discussions for the other species can just replace the operator $%
\hat{a}_{m_{1}}^{\dag }(\hat{a}_{n_{1}})$ with $\hat{b}_{m_{1}}^{\dag }(\hat{%
b}_{n_{1}})$:

\begin{eqnarray}
L_{x} &=&\frac{1}{\sqrt{2}}\left( \hat{a}_{1}^{\dag }\text{ }\hat{a}%
_{0}^{\dag }\text{ }\hat{a}_{-1}^{\dag }\right) \left(
\begin{array}{lll}
0 & 1 & 0 \\
1 & 0 & 1 \\
0 & 1 & 0%
\end{array}%
\right) \left(
\begin{array}{l}
\hat{a}_{1} \\
\hat{a}_{0} \\
\hat{a}_{-1}%
\end{array}%
\right)  \notag \\
&=&\frac{1}{\sqrt{2}}(\hat{a}_{1}^{\dag }\hat{a}_{0}+\hat{a}_{0}^{\dag }\hat{%
a}_{1}+\hat{a}_{0}^{\dag }\hat{a}_{-1}+\hat{a}_{-1}^{\dag }\hat{a}_{0})
\end{eqnarray}

\begin{eqnarray}
L_{y} &=&\frac{i}{\sqrt{2}}(\hat{a}_{1}^{\dag }\text{ }\hat{a}_{0}^{\dag }%
\text{ }\hat{a}_{-1}^{\dag })\left(
\begin{array}{lll}
0 & -i & 0 \\
i & 0 & -i \\
0 & i & 0%
\end{array}%
\right) \left(
\begin{array}{l}
\hat{a}_{1} \\
\hat{a}_{0} \\
\hat{a}_{-1}%
\end{array}%
\right)  \notag \\
&=&\frac{-i}{\sqrt{2}}(\hat{a}_{1}^{\dag }\hat{a}_{0}+\hat{a}_{0}^{\dag }%
\hat{a}_{-1}-\hat{a}_{0}^{\dag }\hat{a}_{1}-\hat{a}_{-1}^{\dag }\hat{a}_{0})
\end{eqnarray}

\begin{eqnarray}
L_{z} &=&(\hat{a}_{1}^{\dag }\text{ }\hat{a}_{0}^{\dag }\text{ }\hat{a}%
_{-1}^{\dag })\left(
\begin{array}{lll}
1 & 0 & 0 \\
0 & 0 & 0 \\
0 & 0 & -1%
\end{array}%
\right) \left(
\begin{array}{l}
\hat{a}_{1} \\
\hat{a}_{0} \\
\hat{a}_{-1}%
\end{array}%
\right)  \notag \\
&=&\hat{a}_{1}^{\dag }\hat{a}_{1}-\hat{a}_{-1}^{\dag }\hat{a}_{-1}
\end{eqnarray}%
The Hamiltonians represented by the creation and annihilation operator are:
\begin{align}
H_{1}& =\beta _{1}\left[ \hat{n}_{1}(\hat{n}_{1}-1)+\hat{n}_{-1}(\hat{n}%
_{-1}-1)\right]  \notag \\
& +\beta _{1}\left[ -2\hat{n}_{1}\hat{n}_{-1}+2\hat{n}_{1}\hat{n}_{0}+2\hat{n%
}_{0}\hat{n}_{-1}\right]  \notag \\
& +\beta _{1}(\hat{a}_{1}^{\dag }\hat{a}_{-1}^{\dag }\hat{a}_{0}\hat{a}_{0}+%
\hat{a}_{0}^{\dag }\hat{a}_{0}^{\dag }\hat{a}_{1}\hat{a}_{-1}),
\end{align}%
\begin{align}
H_{2}& =\beta _{2}\left[ \hat{n}_{1}^{\prime }(\hat{n}_{1}^{\prime }-1)+\hat{%
n}_{-1}^{\prime }(\hat{n}_{-1}^{\prime }-1)\right]  \notag \\
& +\beta _{2}\left[ -2\hat{n}_{1}^{\prime }\hat{n}_{-1}^{\prime }+2\hat{n}%
_{1}^{\prime }\hat{n}_{0}^{\prime }+2\hat{n}_{0}^{\prime }\hat{n}%
_{-1}^{\prime }\right]  \notag \\
& +\beta _{2}(\hat{b}_{1}^{\dag }\hat{b}_{-1}^{\dag }\hat{b}_{0}\hat{b}_{0}+%
\hat{b}_{0}^{\dag }\hat{b}_{0}^{\dag }\hat{b}_{1}\hat{b}_{-1}),
\end{align}%
Compare to the only one spin-exchange process in single spinor BEC: ($a_{0}^{\dag }a_{0}^{\dag
}a_{1}a_{-1}+h.c.$), there are five heteronuclear spin-exchange processes
inside the inter-species part $H_{12}$,
\begin{align}
H_{12}& =\beta (\hat{n}_{1}\hat{n}_{1}^{\prime }+\hat{n}_{-1}\hat{n}%
_{-1}^{\prime }-\hat{n}_{1}\hat{n}_{-1}^{\prime }-\hat{n}_{-1}\hat{n}%
_{1}^{\prime })  \notag \\
& +\gamma (\hat{n}_{0}\hat{n}_{0}^{\prime }+\hat{n}_{1}^{\prime }+\hat{n}%
_{-1}\hat{n}_{1}^{\prime })  \notag \\
& +\beta (\hat{a}_{-1}^{\dag }\hat{a}_{0}\hat{b}_{0}^{\dag }\hat{b}_{-1}+%
\hat{a}_{0}^{\dag }\hat{a}_{-1}\hat{b}_{-1}^{\dag }\hat{b}_{0})  \notag \\
& +\beta (\hat{a}_{1}^{\dag }\hat{a}_{0}\hat{b}_{0}^{\dag }\hat{b}_{1}+\hat{a%
}_{0}^{\dag }\hat{a}_{1}\hat{b}_{1}^{\dag }\hat{b}_{0})  \notag \\
& +(\beta -\gamma )(\hat{a}_{-1}^{\dag }\hat{a}_{0}\hat{b}_{1}^{\dag }\hat{b}%
_{0}+\hat{a}_{0}^{\dag }\hat{a}_{-1}\hat{b}_{0}^{\dag }\hat{b}_{1})  \notag
\\
& +(\beta -\gamma )(\hat{a}_{1}^{\dag }\hat{a}_{0}\hat{b}_{-1}^{\dag }\hat{b}%
_{0}+\hat{a}_{0}^{\dag }\hat{a}_{1}\hat{b}_{0}^{\dag }\hat{b}_{-1})  \notag
\\
& +\gamma (\hat{a}_{1}^{\dag }\hat{a}_{-1}\hat{b}_{-1}^{\dag }\hat{b}_{1}+%
\hat{a}_{-1}^{\dag }\hat{a}_{1}\hat{b}_{1}^{\dag }\hat{b}_{-1}),  \label{A6}
\end{align}

We choose the following set of state vectors as the basis ( or so called
Schwinger presentation), where%
\begin{equation}
\hat{L}_{a}^{+}=\hat{a}_{0}^{\dag }\hat{a}_{-1},\hat{L}_{a}^{-}=\hat{a}%
_{-1}^{\dag }\hat{a}_{0},\hat{L}_{a}^{z}=\frac{1}{2}(\hat{n}_{0}-\hat{n}%
_{-1})
\end{equation}%
\begin{equation}
\hat{L}_{b}^{+}=\hat{b}_{1}^{\dag }\hat{b}_{0},\hat{L}_{a}^{-}=\hat{b}%
_{0}^{\dag }\hat{b}_{1},\hat{L}_{b}^{z}=\frac{1}{2}(\hat{n}_{1}^{\prime }-%
\hat{n}_{0}^{\prime })
\end{equation}%
\newline
and we have,%
\begin{align}
\hat{L}_{a}^{+}\left\vert m_{1},m_{2}\right\rangle & =j_{a}^{+}\left\vert
m_{1}+1,m_{2}\right\rangle , \\
\hat{L}_{a}^{-}\left\vert m_{1},m_{2}\right\rangle & =j_{a}^{-}\left\vert
m_{1}-1,m_{2}\right\rangle , \\
\hat{L}_{b}^{+}\left\vert m_{1},m_{2}\right\rangle & =j_{b}^{+}\left\vert
m_{1},m_{2}+1\right\rangle , \\
\hat{L}_{b}^{-}\left\vert m_{1},m_{2}\right\rangle & =j_{b}^{+}\left\vert
m_{1},m_{2}-1\right\rangle , \\
\hat{L}_{a}^{z}\left\vert m_{1},m_{2}\right\rangle & =m_{1}\left\vert
m_{1},m_{2}\right\rangle , \\
\hat{L}_{b}^{z}\left\vert m_{1},m_{2}\right\rangle & =m_{2}\left\vert
m_{1},m_{2}\right\rangle
\end{align}%
\begin{align}
j_{a}^{\pm }& =\sqrt{(\frac{N_{a}}{2}\pm m_{1}+1)(\frac{N_{a}}{2}\mp m_{1})}
\\
j_{b}^{\pm }& =\sqrt{(\frac{N_{b}}{2}\pm m_{2}+1)(\frac{N_{b}}{2}\mp m_{2})}.
\end{align}

\section{The coherent state}

\label{sec:conherent}

In this section, we give a brief review of quantum ground states of spin-1
BEC. A scalar BEC consisting of $N$ identical bosons can be described as a
Fock state $\left\vert \Psi \right\rangle =(N!)^{-1/2}(\hat{a}^{\dag
})^{N}\left\vert 0\right\rangle $. If the internal degrees of freedom are
considered, we can generalize a Fock state into a coherent state, if
letting:
\begin{equation}
\hat{a}^{\dag }\rightarrow \hat{\Gamma}_{c}^{\dag }=\epsilon _{1}\hat{a}%
_{1}^{\dag }+\epsilon _{0}\hat{a}_{0}^{\dag }+\epsilon _{-1}\hat{a}%
_{-1}^{\dag },
\end{equation}%
with ${\sum_{m=1,0,-1}}\left\vert \epsilon _{m}\right\vert ^{2}=1.$ In the
view of the single-particle concept, $\hat{\Gamma}_{c}^{\dag }$ means
creating one boson in an internal superposition state and the coefficients $%
\epsilon _{m}$ determine the orientation of the internal spin:
\begin{eqnarray}
\epsilon _{1} &=&\exp (-i\varphi )\cos ^{2}(\frac{\theta }{2}),  \notag \\
\epsilon _{0} &=&\sqrt{2}\cos \frac{\theta }{2}\sin \frac{\theta }{2},
\notag \\
\epsilon _{-1} &=&\exp (-i\varphi )\sin ^{2}(\frac{\beta }{2}).
\end{eqnarray}%
As particles are not related to each other in the CSS, we can apply the
direct product $N$ times to get sufficient particle numbers. In the spin-1
system, the CSS is described as:
\begin{equation}
\left\vert \theta ,\varphi \right\rangle =\frac{1}{\sqrt{N!}}(\hat{\Gamma}%
_{c}^{\dag })^{N}\left\vert 0,0,0\right\rangle
\end{equation}%
with all the spins pointing into a definite direction. In particular, if we choose the total spin pointing to the
positive direction of the x-axis ($\theta =\frac{\pi }{2},\varphi =0$), the
CSS reads:
\begin{equation}
\left\vert \frac{\pi }{2},0\right\rangle =\frac{1}{\sqrt{N!}}\left( \frac{1}{%
2}\hat{a}_{1}^{\dag }+\frac{1}{\sqrt{2}}\hat{a}_{0}^{\dag }+\frac{1}{2}\hat{a%
}_{-1}^{\dag }\right) ^{N}\left\vert 0,0,0\right\rangle
\end{equation}

Similarly, the pseudospin-1/2 CSS is described as:
\begin{equation}
\left\vert \theta ,\varphi \right\rangle ^{\left( \frac{1}{2}\right) }=\frac{%
1}{\sqrt{N!}}\left( \cos (\frac{\theta }{2})\hat{a}_{1}^{\dag }+e^{i\varphi
}\sin (\frac{\theta }{2})\hat{a}_{2}^{\dag }\right) ^{N}\left\vert
0,0\right\rangle .
\end{equation}

\section{Two-site Hubbard model}

\label{sec:Hubbard}

A brief review of quantum ground states of the two site Boson-Hubbard model
is necessary. The Hamiltonian is
\begin{equation}
H=-t(a^{\dag }b+b^{\dag }a)+U[n_{a}(n_{a}-1)+n_{b}(n_{b}-1)]
\end{equation}%
with $a$ and $b$ denote the left and right site respectively. The hopping
parameter is t and interaction is U. The Hilbert space is expanded by Fock state $%
\left\vert n_{a},n_{b}\right\rangle $, or rewritten as $\left\vert
m\right\rangle $:%
\begin{equation}
\left\vert m\right\rangle =\frac{(\hat{a}^{\dag })^{\frac{N}{2}+m}}{\sqrt{(%
\frac{N}{2}+m_{1})!}}\frac{(\hat{b}^{\dag })^{\frac{N}{2}-m}}{\sqrt{(\frac{N%
}{2}-m_{1})!}}\left\vert 0\right\rangle.
\end{equation}
Rewriting operators to be $a_{\pm}=(a^{\dag}\pm b^{\dag})/\sqrt{2}$, the coherent ground state of Hubbard model is
\begin{eqnarray*}
\left\vert G\right\rangle &=&\frac{1}{\sqrt{N!}}(a_{+}^{\dag})^{N}\left\vert
0\right\rangle =\frac{1}{\sqrt{2^{N}N!}}(a^{\dag }+b^{\dag })^{N}\left\vert
0\right\rangle \\
&=&\sum\nolimits_{m=-\frac{N}{2}}^{\frac{N}{2}}\frac{C_{N}^{\frac{N}{2}+m}}{
\sqrt{2^{N}N!}}(a^{\dag })^{\frac{N}{2}+m}(b^{\dag })^{\frac{N}{2}%
-m}\left\vert 0\right\rangle,
\end{eqnarray*}
which has a gaussian amplitude distribution.


\begin{thebibliography}{99}
\bibitem{Quantum entanglement} R. Horodecki, P. Horodecki, M. Horodecki, and
K. Horodecki, Quantum entanglement, Rev. Mod. Phys. 81, 865 (2009).

\bibitem{ref2} O. G\"{u}hne and G. T\'{o}th, Entanglement detection, Phys. Rep. 474, 1 (2009).

\bibitem{ref3} K. Kim, M.-S. Chang, S. Korenblit, R. Islam, E. Edwards, J.
Freericks, G.-D. Lin, L.-M. Duan, and C. Monroe, Quantum simulation of frustrated Ising spins with trapped ions,
Nature (London) 465, 590
(2010).

\bibitem{ref4} J. Simon, W. S. Bakr, R. Ma, M. E. Tai, P. M. Preiss, and M.
Greiner, Quantum simulation of antiferromagnetic spin chains in an optical lattice, Nature (London) 472, 307 (2011).

\bibitem{ref5} V. Giovannetti, S. Lloyd, and L. Maccone, Advances in quantum metrology, Nat. Photonics 5,
222 (2011).

\bibitem{ref6} H. Aasi et al., Enhanced sensitivity of the LIGO gravitational wave detector by using squeezed states of light, Nat. Photonics 7, 613 (2013).

\bibitem{ref7} M. A. Taylor, J. Janousek, V. Daria, J. Knittel, B. Hage,
H.-A. Bachor, and W. P. Bowen, Subdiffraction-Limited Quantum Imaging within a Living Cell, Phys. Rev. X 4, 011017 (2014).

\bibitem{GHZ1} M. J. Holland and K. Burnett, Interferometric detection of optical phase shifts at the Heisenberg limit, Phys. Rev. Lett. 71, 1355
(1993).

\bibitem{GHZ2} A. S. Sorensen and K. Molmer, Entanglement and Extreme Spin Squeezing, Phys. Rev. Lett. 86, 4431
(2001).

\bibitem{GHZ3} K. S. Choi, A. Goban, S. B. Papp, S. J. van Enk, and H. J.
Kimble, Entanglement of spin waves among four quantum memories, Nature (London) 468, 412 (2010).

\bibitem{GHZ4} C. Gross, H. Strobel, E. Nicklas, T. Zibold, N. Bar-Gill, G.
Kurizki and M. K. Oberthaler, Atomic homodyne detection of continuous-variable entangled twin-atom states, Nature 480, 219 (2011).

\bibitem{GHZ5} B. L\"{u}ke, M. Scherer, J. Kruse, L. Pezz\'{e}, F.
Deuretzbacher, P. Hyllus, O. Topic, J. Peise, W. Ertmer, J. Arlt, L. Santos,
A. Smerzi, C. Klempt, Twin Matter Waves for Interferometry Beyond the Classical Limit, Science 334, 773 (2011).

\bibitem{GHZ6} Z. Zhang and L.-M. Duan, Generation of Massive Entanglement through an Adiabatic Quantum Phase Transition in a Spinor Condensate,
Phys. Rev. Lett. 111, 180401 (2013).

\bibitem{science2019_1} A. Omran, H. Levine, A. Keesling, G. Semeghini, T. T. Wang,
S. Ebadi, H. Bernien, A. S. Zibrov, H. Pichler, S. Choi, J. Cui,
M. Rossignolo, P. Rembold, S. Montangero, T. Calarco, M.
Endres, M. Greiner, V. Vuletic, and M. D. Lukin, Generation
and manipulation of Schrodinger cat states in Rydberg atom
arrays, Science 365, 570 (2019).

\bibitem{science2019_2} C. Song, K. Xu, H. Li, Y.-R. Zhang, X. Zhang, W. Liu, Q. Guo,
Z. Wang, W. Ren, J. Hao, H. Feng, H. Fan, D. Zheng, D.-W.
Wang, H. Wang, and S.-Y. Zhu, Generation of multicomponent
atomic Schrodinger cat states of up to 20 qubits, Science 365,
574 (2019).

\bibitem{2021npj} Zhao, Y., Zhang, R., Chen, W. et al. Creation of
Greenberger-Horne-Zeilinger states with thousands of atoms by entanglement
amplification. npj Quantum Inf 7, 24 (2021).

\bibitem{TLH} T.-L. Ho, Spinor Bose Condensates in Optical Traps, Phys. Rev. Lett. \textbf{81}, 742 (1998).

\bibitem{Ohmi} T. Ohmi and K. Machida, Bose-Einstein Condensation with Internal Degrees
of Freedom in Alkali Atom Gases, J. Phys. Soc. Jpn. \textbf{67}, 1822 (1998).

\bibitem{Law} C. K. Law, H. Pu, and N. P. Bigelow, Quantum Spins Mixing in Spinor Bose-Einstein Condensates, Phys. Rev. Lett. \textbf{%
81}, 5257 (1998).

\bibitem{Yi} S. Yi, \"{O}. E. M\"{u}stecaplioglu, C. P. Sun, and L. You, Single-mode approximation in a spinor-1 atomic condensate,
Phys. Rev. A \textbf{66}, 011601(R) (2002).

\bibitem{MSChang} M.-S. Chang, Q. S. Qin, W. X. Zhang, L. You, and M. S.
Chapman, Coherent spinor dynamics in a spin-1 Bose condensate, Nat. Phys. \textbf{1}, 111 (2005).

\bibitem{spinorfull} D. M. Stamper-Kurn and M. Ueda, Spinor Bose gases: Symmetries, magnetism, and quantum dynamics, Rev. Mod. Phys. 85,
1191 (2013).

\bibitem{L.M.Duan} A. S$\emptyset $ensen, L.-M. Duan, J. I. Cirac, and P.
Zoller, Many-particle entanglement with Bose¨CEinstein condensates, Nature (London) 409, 63 (2001).

\bibitem{squeezing1} C. Gross, T. Zibold, E. Nicklas, J. Est\`{e}ve, and M.
K. Oberthaler, Atomic homodyne detection of continuous-variable entangled twin-atom states, Nature (London) 464, 1165 (2010).

\bibitem{squeezing2} M.F. Riedel, P. B\"{o}hi, Y. Li, T.W. H\"{a}nsch, A.
Sinatra, and P. Treutlein, Atom-chip-based generation of entanglement for quantum metrology, Nature (London) 464, 1170 (2010).

\bibitem{nematic} C.D.Hamley, C. S. Gerving, T. M. Hoang, E. M. Bookjans,
and M. S. Chapman, Spin-nematic squeezed vacuum in a quantum gas, Nat. Phys. 8, 305 (2012).

\bibitem{Dicke Experiment1} K. S. Choi, A. Goban, S. B. Papp, S. J. van Enk,
and H. J. Kimble, Nature (London) 468, 412 (2010).

\bibitem{Dicke Experiment2} C. Gross, H. Strobel, E. Nicklas, T. Zibold, N.
Bar-Gill, G. Kurizki and M. K. Oberthaler, Nature 480, 219 (2011).

\bibitem{Dicke Experiment3} B. L\"{u}ke, M. Scherer, J. Kruse, L. Pezz\'{e},
F. Deuretzbacher, P. Hyllus, O. Topic, J. Peise, W. Ertmer, J. Arlt, L.
Santos, A. Smerzi, C. Klempt, Science 334, 773 (2011).

\bibitem{twinfock} M. J. Holland and K. Burnett, Phys. Rev. Lett. 71, 1355
(1993).
\bibitem{indicator} A. S. Sorensen and K. Molmer, Entanglement and Extreme Spin Squeezing, Phys. Rev. Lett. 86, 4431
(2001).

\bibitem{LiYouTwinfock} Xin-Yu Luo, Yi-Quan Zou, Ling-Na Wu, Qi Liu, Ming-Fei Han, Meng Khoon Tey, and Li You, Deterministic entanglement
generation from driving through quantum phase transitions, Science 355, 620-623 (2017).


\bibitem{mixture1} M. Luo, Z. Li, and C. Bao, Bose-Einstein condensate of a mixture of two species of spin-1 atoms, Phys. Rev. A 75, 043609
(2007).

\bibitem{mixture2} Zhi-fang Xu, Jie Zhang, Yunbo Zhang, and Li You, Quantum states of a binary mixture of spinor Bose-Einstein condensates, Phys. Rev. A 81, 033603
(2010).

\bibitem{mixture3} Jie Zhang, Zhi-Fang Xu, Li You, and Yunbo Zhang, Atomic-number fluctuations in a mixture of condensates, Phys. Rev. A
82, 013625 (2010).

\bibitem{mixture4} Jie Zhang, Tiantian Li, and Yunbo Zhang, Interspecies singlet pairing in a mixture of two spin-1 Bose condensates, Phys. Rev. A 83, 023614
(2011).

\bibitem{mixture5} Yu Shi, Ground states of a mixture of two species of spinor Bose gases with interspecies spin exchange, Phys. Rev. A 82, 023603 (2010).

\bibitem{mixture6} Jie Zhang, Xue Hou, Bin Chen, and Yunbo Zhang, Fragmentation of a spin-1 mixture in a magnetic field, Phys. Rev.
A 91, 013628 (2015).

\bibitem{mixture7} F. D.Wang, D. Z. Xiong, X. K. Li, D. J.Wang, and E.
Tiemann, Observation of Feshbach resonances between ultracold Na and Rb atoms, Phys. Rev. A 87, 050702(R) (2013).

\bibitem{mixture8} Xiaoke Li, Bing Zhu, Xiaodong He, Fudong Wang, Mingyang
Guo, Zhi-Fang Xu, Shizhong Zhang, and Dajun Wang, Coherent Heteronuclear Spin Dynamics in an Ultracold Spinor Mixture, Phys. Rev. Lett. \textbf{%
114}, 255301 (2015).

\bibitem{KRb} A. Burchianti, C. D'Errico, S. Rosi, A. Simoni, M. Modugno, C. Fort and F. Minardi,
Dual-species Bose-Einstein condensate of 41K and 87Rb in a hybrid trap, Phys. Rev. A \textbf{%
98}, 063616 (2018).

\bibitem{KRbdroplet} C. D'Errico, A. Burchianti, M. Prevedelli, L. Salasnich, F. Ancilotto,
M. Modugno, F. Minardi, and C. Fort, Phys. Rev. Research \textbf{%
1}, 033155 (2019).


\bibitem{Hocat} T.-L. Ho and C. V. Ciobanu, The Schrodinger Cat Family in
Attractive Bose Gases and Their Interference, arXiv.0011095v1 (2000).

\bibitem{HoYip} T.-L. Ho and S.-K. Yip, Fragmented and Single Condensate Ground States of Spin-1 Bose Gas, Phys. Rev. Lett. \textbf{84}, 4031
(2000).

\bibitem{n396} J. Stenger, S. Inouye, D. M. Stamper-Kurn, H.-J. Miesner, A.
P. Chikkatur and W. Ketterle, Spin domains in ground-state Bose¨CEinstein condensates, Nature (London) \textbf{396}, 345 (1998).

\bibitem{n443}  L.E. Sadler, J. M. Higbie, S. R. Leslie, M. Vengalattore, and D. M.
Stamper-Kurn, Spontaneous symmetry breaking in a quenched ferromagnetic spinor Bose¨CEinstein condensate, Nature (London) \textbf{443}, 312 (2006).


\end{thebibliography}
\end{document}